\documentclass[aps,reprint,graphicx,groupedaddress,amsmath,amssymb,longbibliography]{revtex4-1}
\usepackage{dcolumn}% Align table columns on decimal point
\usepackage{bm}% bold math
\usepackage{natbib}
\usepackage[usenames,dvipsnames] {color}
\usepackage[table]{xcolor}
\usepackage{graphicx}
\usepackage{epstopdf}
\usepackage{amssymb}
\usepackage{amsmath}
\usepackage{dcolumn}
\usepackage{subfigure}

\DeclareGraphicsExtensions{.eps,.png,.pdf}

\usepackage{hyperref}
\hypersetup{
  colorlinks   = true, %Colours links instead of ugly boxes
  urlcolor     = blue, %Colour for external hyperlinks
  linkcolor    = blue, %Colour of internal links
  citecolor   = blue %Colour of citations
}
\usepackage{setspace}
\usepackage[font=singlespacing,font=footnotesize, justification=RaggedRight]{caption}
\usepackage{enumitem}
\usepackage{fancybox}
\usepackage{comment}
\usepackage{url}
\usepackage{here}
\setlist{nolistsep}
\usepackage{multimedia}
\usepackage{soul}
\setstcolor{red}
\DeclareMathAlphabet      {\mathbf}{OT1}{cmr}{bx}{n}

\newcommand{\printfnsymbol}[1]{%
  \textsuperscript{\@fnsymbol{#1}}%
}
\makeatother

\graphicspath{{./Figures/}}

\begin{document}

\title{Liquid-Hexatic-Solid Phase Transition of a Hard-Core Lattice Gas with Third Neighbor Exclusion}

\author{Shaghayegh Darjani$^{1,3\dag}$}
\author{Joel Koplik$^{2}$}
\email{jkoplik@ccny.cuny.edu}
\author{Sanjoy Banerjee$^{1}$}
\author{Vincent Pauchard $^{1}$}
\affiliation{
	$^1$Energy Institute and Department of Chemical Engineering, City College of New York, New York, NY 10031, USA. \\
	$^2$Benjamin Levich Institute and Department of Physics, City College of New York, NY 10031, USA.\\
	$^3$Benjamin Levich Institute and Department of Chemical Engineering, City College of the City University of New York, New York, New York 10031, USA
	}

\date{\today}

\begin{abstract}
\begin{singlespace}
{The determination of phase behavior and, in particular, the nature of phase transitions in two-dimensional systems is often clouded by finite size effects and by access to the appropriate thermodynamic regime. We address these issues using an alternative route to deriving the equation of state of a two-dimensional hard-core particle system, based on kinetic arguments and the Gibbs adsorption isotherm, by use of the random sequential adsorption with surface diffusion ($RSAD$) model. Insight into coexistence regions and phase transitions is obtained through direct visualization of the system at any fractional surface coverage via local bond orientation order. The analysis of the bond orientation correlation function for each individual configuration confirms that first-order phase transition occurs in a two-step liquid-hexatic-solid transition at high surface coverage. }
\end{singlespace}
\end{abstract}

\maketitle

\section{Introduction}
The phase behavior of the two dimensional systems is a key aspect of many current research areas such as emulsion stability due to particle adsorption at the interface \cite{pauchard2014asphaltene,liu2017mixture}, particle self-assembly into clusters \cite{hou2019phase,fortuna2010hexagonal,weber2008role,gorbunov2013adsorption}, chemisorption on metal surface \cite{bak1985phase,taylor1985two} and melting at an interface \cite{bernard2011two,engel2013hard}. Numerous equations of state ($EOS$) such as the Langmuir \cite{langmuir1918adsorption} and Volmer models have been introduced over the years to describe the adsorption behavior of such systems \cite{rane2015applicability,rane2013interfacial}. The Langmuir model is based on localized adsorption where the adsorbate molecules are smaller than the adsorption sites\cite{foo2010insights}, but in many practical cases the adsorbate is actually larger than the adsorption site. Likewise, the Volmer model, which assumes fully delocalized adsorption (adsorbates much bigger than the adsorption sites), is equally inappropriate except, perhaps, for nanoparticles. Rather than directly prescribing the $EOS$, an alternative is the random sequential adsorption model ($RSA$), which describes the dynamics of the adsorption process by allowing objects to adsorb sequentially onto the open sites of a one or two-dimensional lattice \cite{wang1993collective}. The model has been extensively used in the literature, for example by Manzi et al. \cite{manzi2019simulations} who discuss the adsorption of human serum albumin on the nano-structure of a black silicon surface using $RSA$. Further applications of this model include chemisorption, deposition, layered growth, vibrated granular material and the car-parking problem \cite{lonvcarevic2009adsorption,lonvcarevic2007reversible,talbot2000car,evans1993random,krapivsky1994collective,wang1993locally,vscepanovic2016response,tarjus2004statistical,meineke2001random,schaaf1989surface}. \par

More specifically, in the $RSA$ model molecules or particles are progressively added at random to an initially empty surface with the only restriction that overlap is not allowed, an assumption based physically on short-range electrostatic repulsion. As the coverage increases, the free area left for further adsorption decreases, not only because of the sites occupied by previously adsorbed molecules, but also because vacancies can be too small to allow adsorption without overlap. Without desorption or surface diffusion, adsorption kinetic rapidly slows down and coverage only asymptotically approaches the jamming limit, equivalent to random maximum packing, if the substrate is not pre-patterned. However, it has been experimentally observed \cite{ramsden1992observation} that the relaxation time scale of adsorbed particles, due to their rearrangement on the surface, can be comparable to the deposition time scale. The final configuration is comparable with a dense-packed ordered system. Furthermore, none of the above models is able to explain the ordered layering observed in adsorption of certain materials on the surface \cite{bak1985phase,taylor1985two,ramsden1992observation}.\par
A different approach is taken in the lattice gas model, which uses statistical mechanical reasoning to describe adsorbate configurations on a surface, where an adsorbate molecule may occupy one or a few adsorption sites. Applications of this model include a recent study of photo-excited Rydberg gases by Ji et al. \cite{ji2011two}, where an order-disorder phase transition corresponds to the phase transition on a square lattice with first neighbor exclusion, a study of self-assembly of isophthalic acid on graphite by Lackinger et al. \cite{lackinger2004self}, adsorption of selenium on Nickel surface by Bak et al. \cite{bak1985phase} and chemisorption of oxygen on palladium by Zhang et al. \cite{zhang2007accuracy}. Although the lattice gas model has been studied extensively in the literature, only the single case of a triangular lattice with first neighbor exclusion was solved exactly, by Baxter \cite{baxter2016exactly}.  For all other variants, a number of lattice gas methods have been developed over years based on various approximation methods: the matrix method of Kramer and Wannier \cite{bellemans1967phase,orban1968phase,ree1966phase,runnels1967exact,runnels1966exact,feng2011lattice,rotman2009critical}, the density (or activity) series expansion method \cite{bellemans1967phase,orban1968phase,gaunt1965hard,gaunt1967hard,eisenberg2005first,ushcats2015high,ushcats2016virial} , the generalized Bethe method \cite{cowley1979bethe,burley1960lattice,hansen2005hard}, Monte Carlo simulation \cite{feng2011lattice,binder1980phase,chesnut1971monte,liu2000ordering,fernandes2007monte,nath2014multiple}, the Rushbrooke and Scoins method \cite{rushbrooke1955ising} and fundamental measure theory \cite{lafuente2003density}. Despite all of this effort, the lattice gas model is not able to describe the adsorption isotherm of the system, and instead focuses on the equation of state and the nature of phase transition.\par
In order to combine the advantages of both the RSA and lattice gas models, we previously developed an alternative route to deriving the equation of state of a two-dimensional hard-core particle with first neighbor exclusion, based on kinetic arguments and the Gibbs adsorption isotherm: the $RSAD$ model \cite{darjani2017extracting}. Here one considers a two-dimensional lattice gas in equilibrium with a three-dimensional solution of adsorbate molecules, where the equality of chemical potential throughout the system leads to:
\begin{equation}
d\Pi = kT\,\frac{\Theta}{A_a}\,d\ln C
\label{1}
\end{equation}
where $A_a$ is the interfacial area covered by a single adsorbate molecule, $\theta$ is the fractional surface coverage and $C$ is the concentration of the (three-dimensional) solution. Integrating the above equation gives:
\begin{equation}
\int\limits_0^\Theta  {\frac{\Theta }{C}} \,\frac{{\partial C}}{{\partial \Theta }}\,d\Theta  = \frac{{{A_a}}}{{kT}}\,\Pi 
\label{2}
\end{equation}
From which we see that knowledge of the adsorption isotherm, the relationship between $C(\Theta)$, bulk concentration, and fractional coverage, enables one to calculate the equation of state $\Pi(\Theta)$.\par 
The adsorption isotherm, in turn, can be obtained through kinetic arguments. At equilibrium the rates of adsorption and desorption of molecules are equal:
\begin{equation}
K_a\,C(1-\beta(\Theta)) = K_d\,\Theta
\label{3}
\end{equation}
where $K_a$ and $K_d$ are the adsorption and desorption rate constants, respectively, and $\beta (\Theta )$ is the ``blocking function€™", the fraction of the surface area which is excluded from further adsorption by already adsorbed molecules. Solving for $C$ and inserting the result into the integral version of the Gibbs adsorption isotherm yields:
\begin{equation}
\int\limits_0^\Theta  {(1 - \beta (} \Theta ))
\frac{\partial }{{\partial \Theta }}
\left[\frac{\Theta }{1 - \beta (\Theta )}\right]\,d\Theta  
= \frac{{{A_a}}}{{kT}}\,\Pi 
\label{5}
\end{equation}
Thus, the blocking function is the only information needed to calculate the equation of state, and we have shown previously \cite{darjani2017extracting} that for lattice gases the blocking function can be easily extracted from $RSAD$ model simulations.  \par
In the $RSAD$ model, where surface diffusion is introduced in parallel with adsorption, vacancies large enough to adsorb a further particle are both created and destroyed. When diffusion is sufficiently rapid, the size distribution of vacancies no longer depends on the history of adsorption (the positions where the adsorbates first arrived on the substrate) but only on the fractional surface coverage. One of the advantages of using the $RSAD$ model is in locating the equilibrium state, which assures us that enough thermalization is present to reach the equilibrium state. Note that in this model the potential energy is effectively infinite for particle overlap, due to the repulsive interaction, which restricts the occupancy of neighbors, and is zero otherwise. The system can therefore be considered as athermal \cite{rotman2009critical,fernandes2007monte,darjani2017extracting}. Our results show that the $RSAD$ model can be used as an equilibrium model and our equation of state, the nature of our phase transition and the phase transition coverage are in excellent agreement with the only model with an exact solution in the literature \cite{baxter2016exactly}. From the definition of the adsorption rate, used above to define adsorption equilibrium, the blocking function can be extracted from the numerical simulations through the derivative of surface coverage with respect to the number of attempts:
\begin{equation}
\frac{{\partial N}}{{\partial n}} = 1 - \beta (\Theta )
\label{6}
\end{equation}
Here $N$ is the number of adsorbed molecules, $n$ is the number of attempts and t, defined by $nA_a/A$=$K_aC/t$, is an adimensional time defined via the adsorption rate. The latter definition will be used in practice by equating the blocking function to the rebuttal rate of adsorption attempts. Ushcats et al. \cite{ushcats2016virial} used an alternative method based on the power of activity at low and high densities, which shows the importance of accounting for the holes in deriving the equation of state of lattice gases. Based on their method \cite{ushcats2016virial,ushcats2015high}, hole-particle symmetry, the total interaction energy is directly related to the interaction of holes at any specific configuration.  \par
In this paper we study the phase behavior of hard-core molecules with third neighbor exclusion on a triangular lattice and compare our results with those of Orban et al. \cite{orban1968phase} who studied this model previously. Hard-core molecules with extended exclusion ranges are studied extensively in the literature but mainly on a square lattice \cite{feng2011lattice,rotman2009critical,fernandes2007equation,fernandes2007monte,lafuente2003phase,ramola2012high,mandal2017estimating,nath2014multiple,nath2016high}. In some experiments it is observed that lateral interactions of adsorbed particles on solid surfaces (chemisorption) follow the extended exclusion range, which is important in surface science as ordering affects the surface functionality \cite{bak1985phase,taylor1985two,zhang2007accuracy,liu2004lattice}. Increasing the exclusion range could also correspond to a smaller lattice site which becomes equivalent to the continuum limit when the exclusion range is significantly large \cite{nath2014multiple,fernandes2007monte,orban1968phase} .\par
Simulation data related to the triangular lattice with third neighbor exclusion is given in the next section, where a first order liquid-hexatic-solid phase transition is obtained at high surface coverage. In the liquid state, positional and orientational correlations of particles decay exponentially, while in solid state they have a quasi-long-range orientational order. Besides these two phases there is an intermediate hexatic phase, where particles have quasi-long range orientational order and exponential positional order. We quantify the ordering using a bond orientation correlation function, ${g_6}(r)$, calculated from the local bond orientation order, $\Psi ({r})$. In a dense system, most of the particles are surrounded by six particles, and the local bond orientation is represented by the sixfold orientation:  
\begin{equation}
\Psi ({r_j}) = \frac{1}{{{N_k}}}\sum\limits_{k = 1}^{{N_k}} {{e^{i6{\theta _{jk}}}}} 
\label{7}
\end{equation}
Where $k$ is the number of the nearest neighbors of particle $j$, $\theta _{jk}$ is the angle between the line joining the centers of mass of particles $j$ and $k$ and a reference axis. The bond orientation correlation function is then defined as an average of the local bond orientation order:
\begin{equation}
{g_6}(r) = \frac{{\left\langle {\sum\limits_{k \ne j}^N {\Psi {{({r_j})}^*}\Psi ({r_k})\delta (r - \left| {{r_j} - {r_k}} \right|)} } \right\rangle }}{{\left\langle {\sum\limits_{k \ne j}^N {\delta (r - \left| {{r_j} - {r_k}} \right|)} } \right\rangle }}
\label{8}
\end{equation}
A power-law decay of ${g_6}(r)$ means that there is a quasi-long-range orientation correlation. For the system in an hexatic phase ${g_6}(r) \propto {r^{ - \eta }}$ where $0 < \eta  < 0.25$. \par
Liquid-solid transitions have been reported extensively in the literature for two-dimensional systems in lattice gas models with extended exclusion ranges. However here our simulation results reveal a liquid-hexatic-solid phase transitions for the triangular lattice cover seven sites, which have the same nature of the phase transition as the melting transition of hard-disk molecules. Although melting is studied extensively in the literature \cite{bernard2011two,engel2013hard}, there is still controversy about the nature of its phase transition that can arise from finite size effect or inefficiency of the system to equilibrate the system. In two-dimensional systems, there are three different scenarios for the melting transition of hard disks in the literature: 
\begin{itemize}
\item {Kosterlitz, Thouless, Halperin, Nelson and Young ($KTHNY$) scenario \cite{young1979melting,nelson1979dislocation,kosterlitz1973ordering}: a two-step continuous phase transition, first from liquid to hexatic phase and then from hexatic to solid phase \cite{mak2006large,peng2010melting},}
\item {Two step transition: first order phase transition from liquid to hexatic phase and then continuous phase transition from hexatic to solid phase \cite{bernard2011two,engel2013hard},}
\item {First order phase transition from liquid to solid phase \cite{alder1962phase}.}
\end{itemize}
We will show in the next section how finite size effects and access to the thermodynamic regimes can bring uncertainty regarding phase behavior of the system; mainly the nature of phase transition where these issues are extensively reported in the literature about the nature of melting transition \cite{bernard2011two,wierschem2011simulation}. 

\section{Simulation details}

\begin{figure}[!htb]
\centering
\includegraphics[width=0.9\columnwidth]{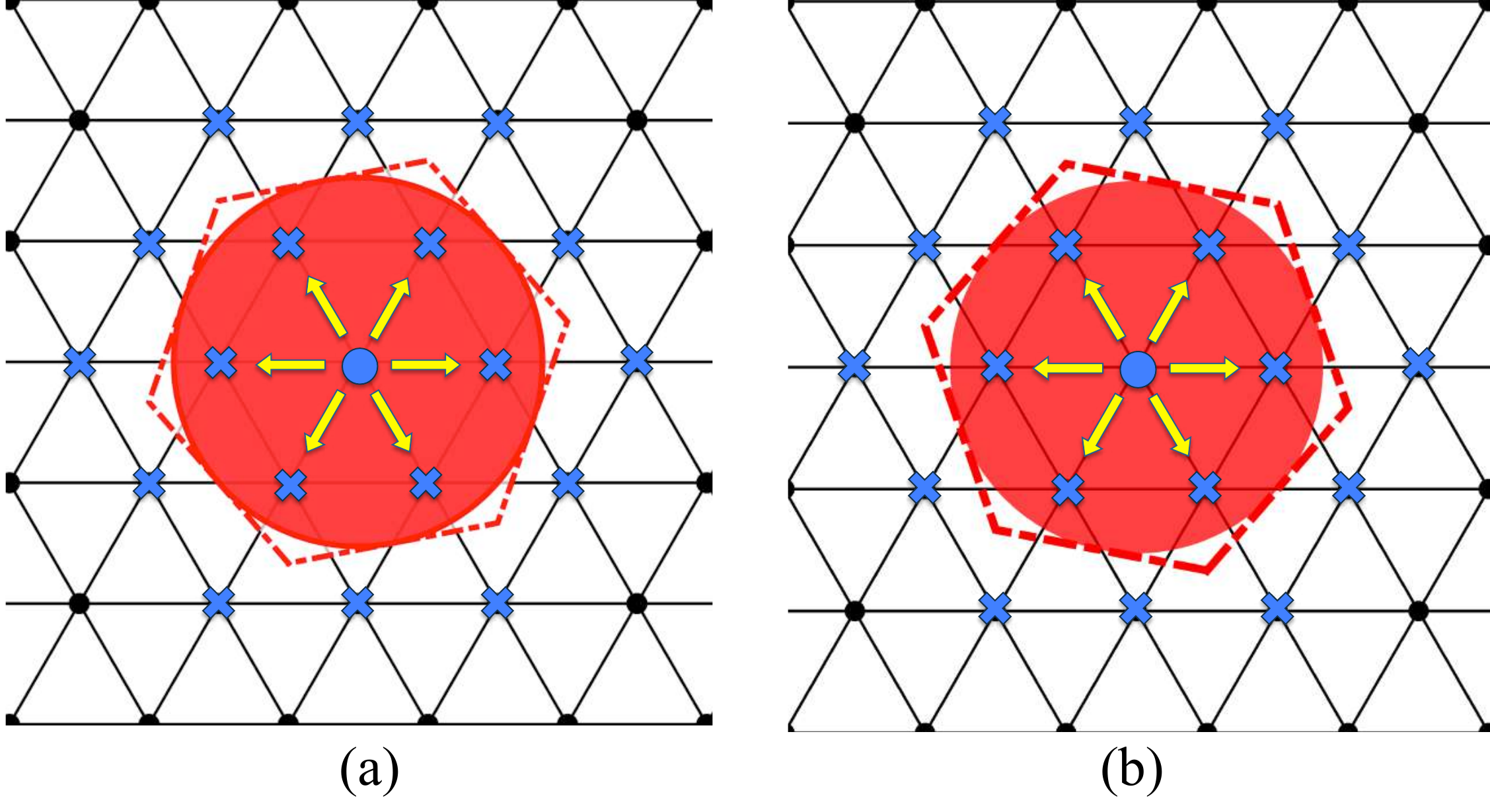}
\caption{Triangular lattice with first, second and third neighbor exclusion where each adsorbate covers $7$ sites. The center of the adsorbate is represented by a circle, arrows indicate possible displacements of particles, and stars represent the sites where the center of other particles are not allowed to adsorb. (a) and (b) show two different configuration at the maximum close packing.}
 \label{fig1}
 \end{figure}

The adsorption of hard-core molecules with a third neighbor exclusion range on the triangular lattice involves the adsorption of molecules covering 7 adsorption sites in the manner as represented by a red circle in Figure.~\ref{fig1}. As it is clear from the hexagon drawn around the adsorbate (red circle) in Figure.~\ref{fig1}(a)-(b), this model could have two different orientations at high coverage in comparison to other adsorbates such as hard-core molecule with first neighbor exclusion  \cite{darjani2017extracting}. Here, we employ two complementary methods as in our previous work  \cite{darjani2017extracting}: an ``adsorption method", which begins from an empty lattice, and a ``desorption method", which begins with a full lattice and progressively decreases coverage. The results are expected to bracket the correct equilibrium equation of state.

In the adsorption method, molecules or particles are progressively added to an initially empty d$\times$d lattice surface where a periodic boundary condition is used to ameliorate finite size effects. The only restriction is that overlap is not allowed; an assumption based physically on short-range electrostatic repulsion. For each adsorption attempt, a random position $(x,y)$ is selected representing the center of mass of the particle. If the selected site and its neighbors are empty, adsorption is accepted. Otherwise it is rejected. Diffusion, the simultaneous movement of particles, is introduced sequentially with a predefined ratio $D$ between the number of diffusion attempt and the adsorption attempt: For $D=3$ each adsorption is followed by 3 diffusion attempts, etc. For each diffusion attempt, a previously adsorbed particle and a direction for the displacement of the particle are selected randomly; an arrow in Figure.~\ref{fig1} illustrates the possible direction. If moving the center of the mass of the particle to the next node along this direction does not infringe the non-overlap condition, diffusion is accepted. Otherwise it is rejected. In the RSAD model, when diffusion is fast enough, the surface layer is at internal equilibrium (even during transient adsorption) and the blocking function can be considered as a state function.\par
For the desorption method the lattice is initially full. For each simulation step, two particles are randomly selected and removed. Then one adsorption attempt and $D$ diffusion attempts are performed following the same procedure as for the adsorption method. The choice of the sequence ($2$ desorption events followed by $1$ adsorption) is arbitrary but answers the need at each time step to decrease coverage and add at least one particle to calculate the blocking function.  \par
For both adsorption and desorption methods, the blocking function is extracted from the success rate of adsorption attempts. $500$ runs are performed, and an ensemble average is used to reduce the noise arising from the numerical calculation of the derivative of the coverage. The blocking function is fitted with a polynomial function before it is used to generate the adsorption isotherm. The latter is inserted into the Gibbs adsorption isotherm equation to obtain the equation of state.\par
In this work we also used a relaxation method in order to track the structure of particles by time and proof our hypothesis about expecting the correct equilibrium equation of state starting from either the adsorption or desorption method. In this method, either the adsorption or desorption method can be used to reach a specific coverage. In the second step, one adsorption attempt and $D$ diffusion attempts are performed following the same procedure as for the adsorption method. In order to keep the coverage constant, if the adsorption attempt is successful, one particle is randomly selected and removed. $1500$ runs are performed to extract the success rate of adsorption attempts or, in other words, the blocking function.\par

\section{Results}
\begin{figure}[!htb]
\centering
\includegraphics[width=1\columnwidth]{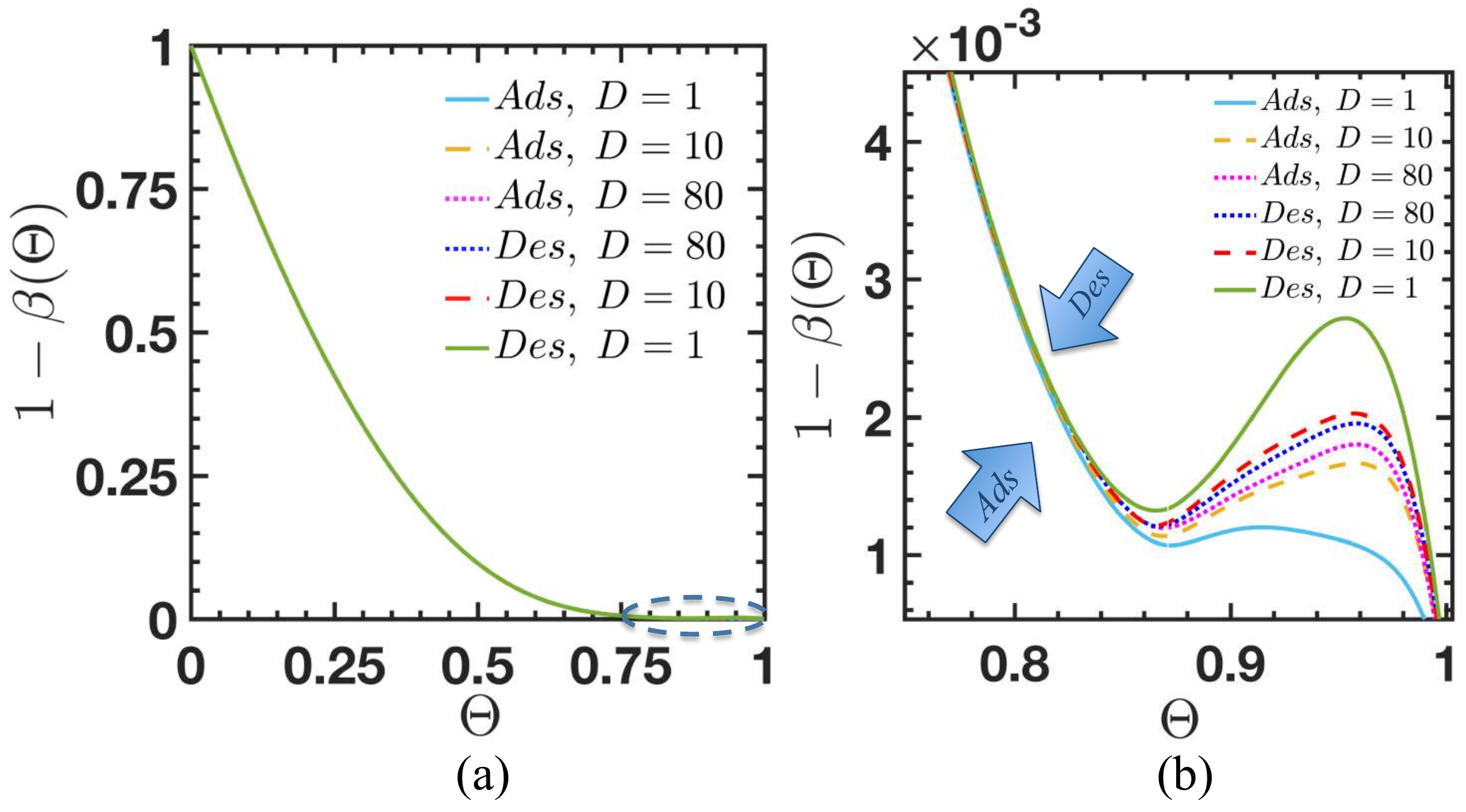}
\caption{ (a) Effect of surface diffusion on adsorption rate versus surface coverage for $d=105$. Ads and Des refer to adsorption and desorption methods, respectively. (b) The inset expands the high-coverage region where sensitivity to surface diffusion appears.}
 \label{fig2}
 \end{figure}

Accessing the thermodynamic regime is an initial step toward studying the phase behavior of the system \cite{bernard2011two,engel2013hard} where some algorithms are insufficient to equilibrate the system in order to study the phase transition of the system in a reliable manner \cite{mak2006large}. The effect of surface diffusion for a lattice size $d=105$ is studied in Figure.~\ref{fig2}. Initially, when the system is diluted, all of the curves regardless of their methods or their surface diffusion overlap in the low surface coverage as presented in Figure.~\ref{fig2}(a). The fluctuation in the phase transition region is much larger than the pure liquid and solid phase, where the difference is maximized in the middle of phase transition as illustrated in Figure\ref{fig2}(b) \cite{alder1962phase}.  As presented in Figure\ref{fig2} (b), at high surface coverage probability of success of adsorbing a new particle initially decreases due to the caging effect. However by increasing the ordering of particles, this caging effect will be diminished in order to maximize the available surface for accepting the new incoming particles. The system reaches the equilibrium state when two curves overlap in the whole range of fractional surface coverage, so accessing the thermodynamic regime will be apparent.  \par
\begin{figure}[!htb]
\centering
\includegraphics[width=0.8\columnwidth]{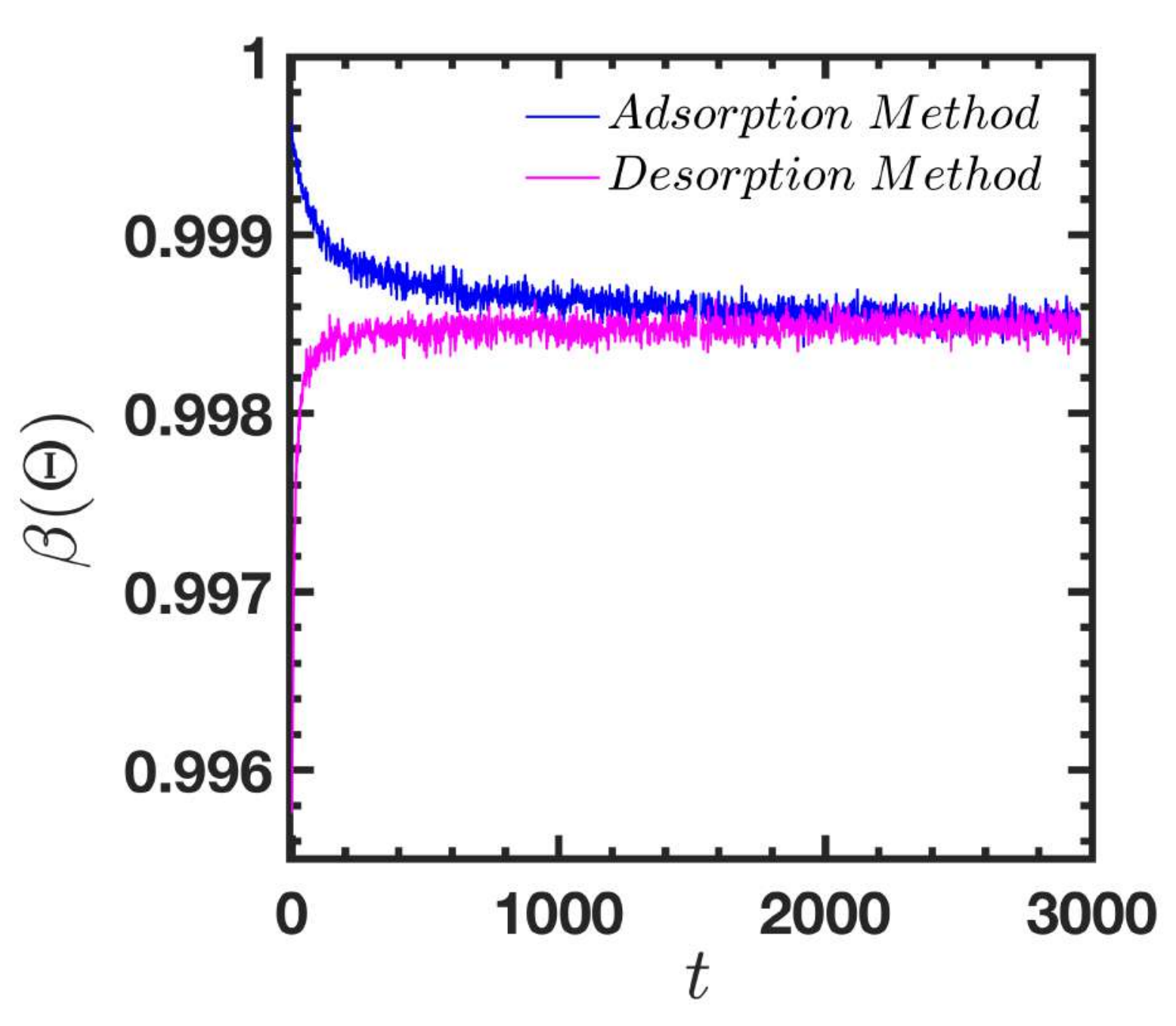}
\caption{Relaxation dynamics of hard core molecule with third neighbor exclusion on triangular lattice for surface coverage of $\theta=0.915$ and $d=196$. $D=0.01$ in both first and second steps. Adsorption and desorption method refer to the initial configuration to reach the surface coverage of $0.915$.}
 \label{fig3}
 \end{figure}

Figure.~\ref{fig3} represents the relaxation of blocking function of hard-core molecules with third neighbor exclusion on a triangular lattice for $D=0.01$, $d=196$ and $\theta=0.915$. This figure confirms that the equilibrium state is a function of a blocking function and does not depend on initial configuration. The blocking function of the desorption method increases in time until it reaches the equilibrium value, whereas the adsorption method shows the opposite trend. Simulation data shows that the desorption method reaches the equilibrium value faster than the adsorption method.\par
 \begin{figure}[!htb]
\centering
\includegraphics[width=1\columnwidth]{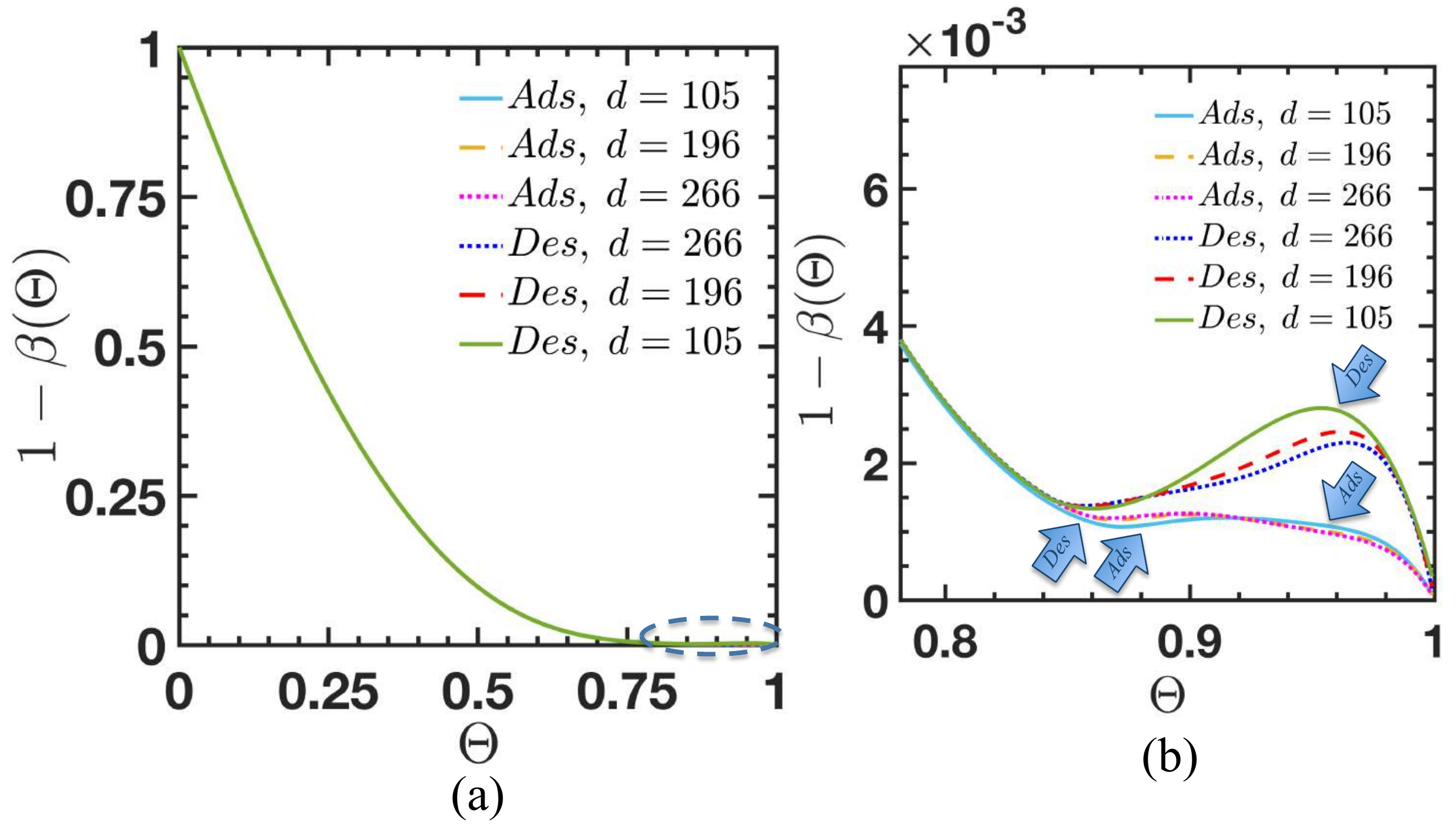}
\caption{(a) Effect of lattice size on adsorption rate versus surface coverage for $D=1$.  Ads and Des refer to adsorption and desorption methods, respectively. (b) the inset expands the high-coverage region where sensitivity to lattice size appears.}
 \label{fig4}
 \end{figure}
\begin{figure}[!htb]
\centering
\includegraphics[width=1\columnwidth]{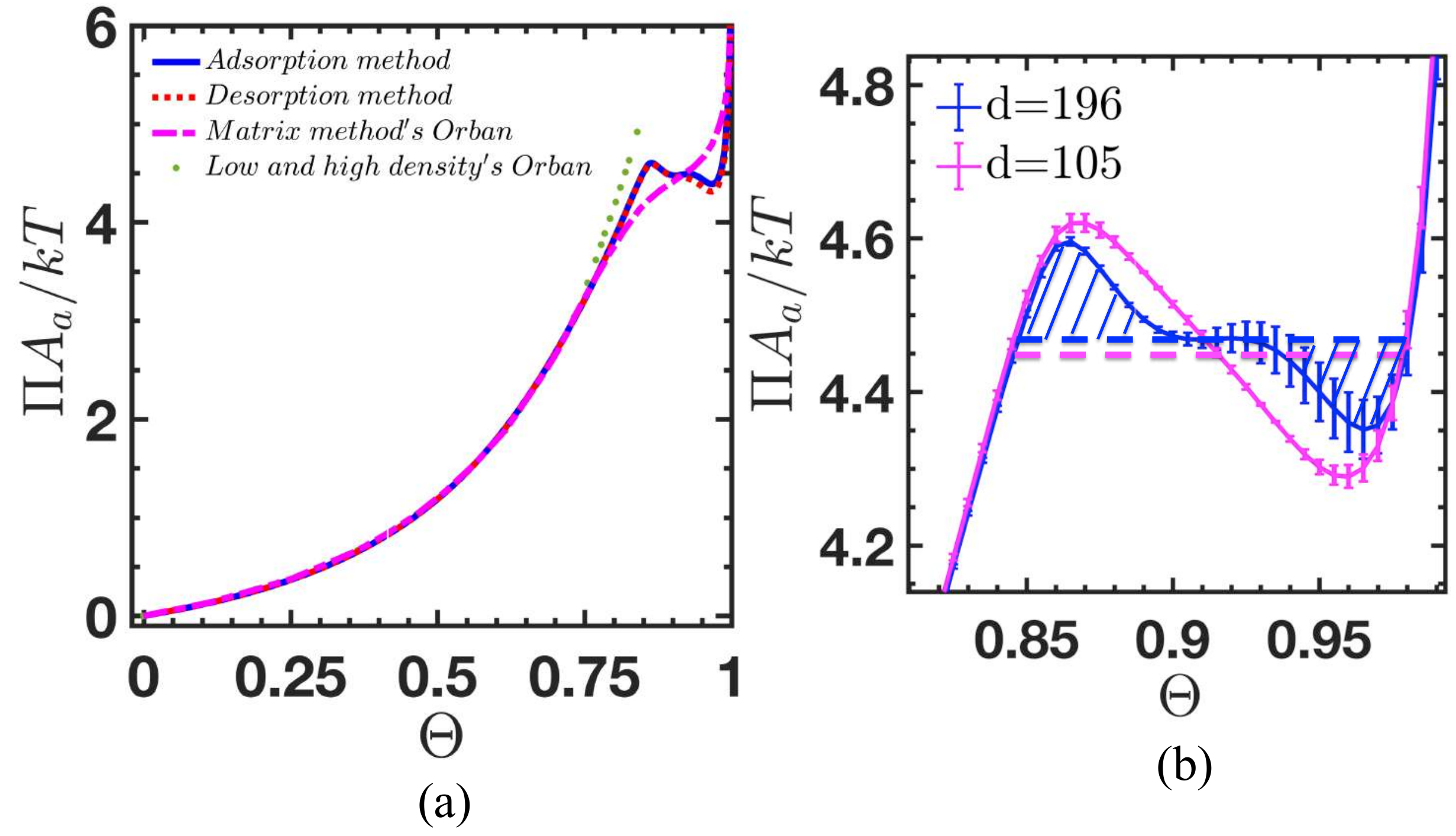}
\caption{ (a) Comparison between our EOS where $d=196$ and $D=80$ with Orban and Bellman (Matrix method and low and high density method) \cite{orban1968phase} for hard core molecules on a triangular lattice covers 7 sites. (b) The inset shows a magnified view of error bar between adsorption and desorption method for ($d=196$,$D=80$) and ($d=105$,$D=80$) in the phase transition region.}
 \label{fig5}
 \end{figure}
 \begin{figure}[!htb]
\centering
\includegraphics[width=0.8\columnwidth]{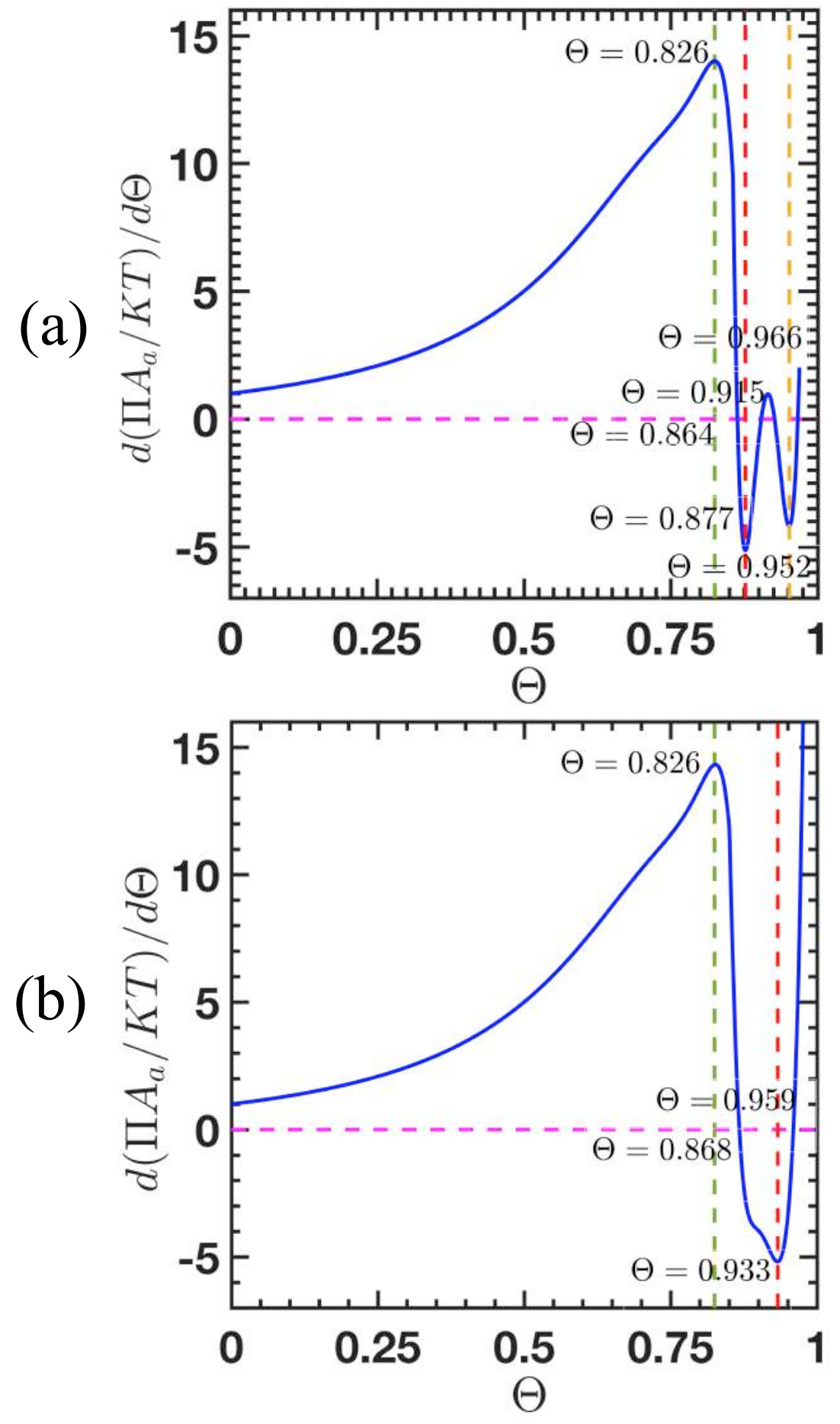}
\caption{ (a) Analysis of phase transition region where $d=196$ and $D=80$, (b) $d=105$ and $D=80$.}
 \label{fig6}
 \end{figure}
   \begin{figure*}[!htb]
\centering
 \includegraphics[width=1\textwidth]{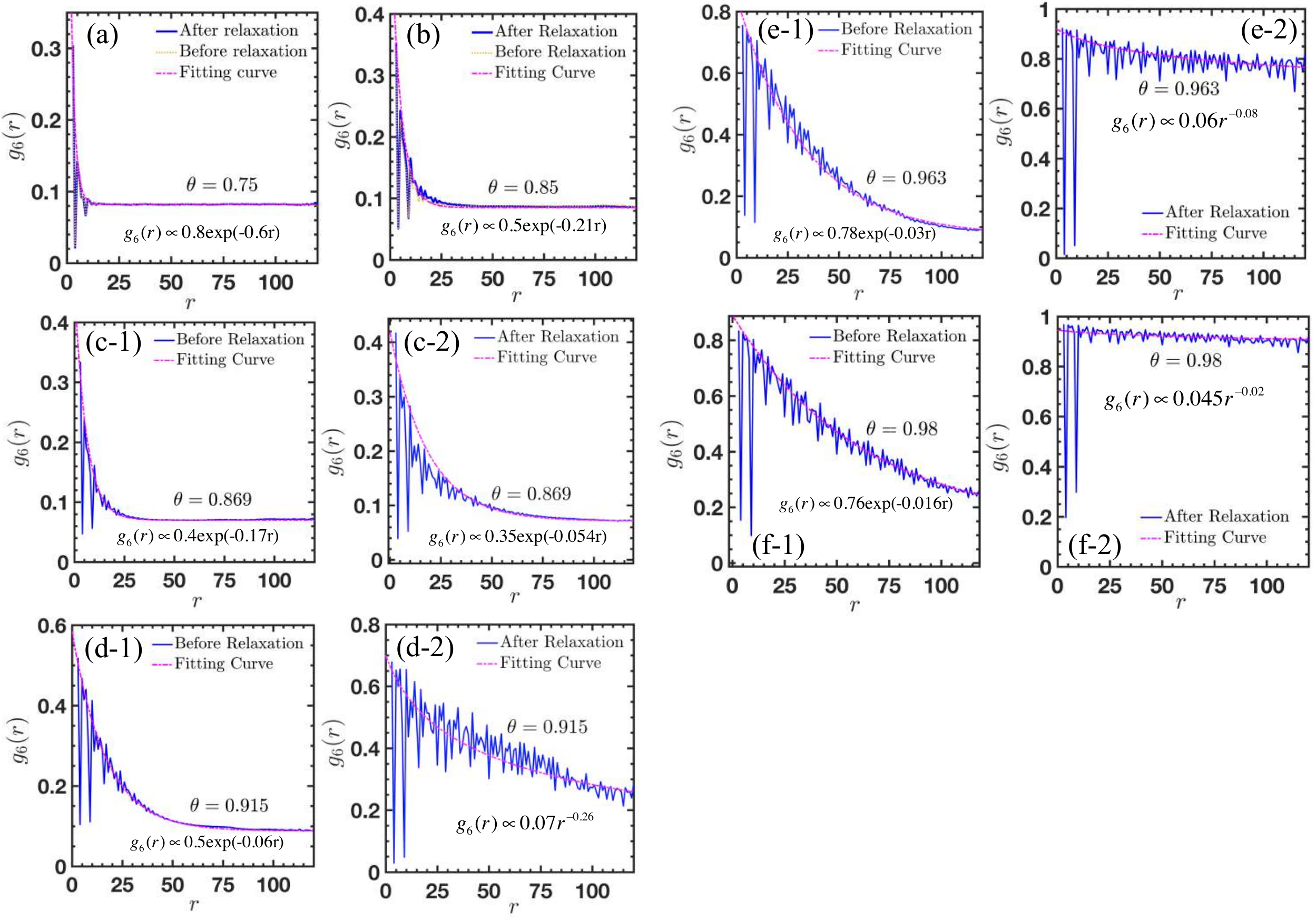}
\caption{Bond orientation correlation function of the hard core molecule with third neighbor exclusion based on relaxation method where $d=196$ and $D=0.01$ in both step of relaxation at different surface coverage (a) $\theta=0.75$, (b) $\theta=0.85$, (c-1) and (c-2) $\theta=0.869$, (d-1) and (d-2) $\theta=0.915$, (e-1) and (e-2) $\theta=0.963$, (f-1) and (f-2) $\theta=0.98$. Before relaxation refer to the configuration obtained from adsorption method at $D=0.01$ in the first step of relaxation method , and after relaxation refer to the equilibrium configuration.}
 \label{fig7}
 \end{figure*}
 \begin{figure*}[!htb]
\centering
\includegraphics[width=1\textwidth]{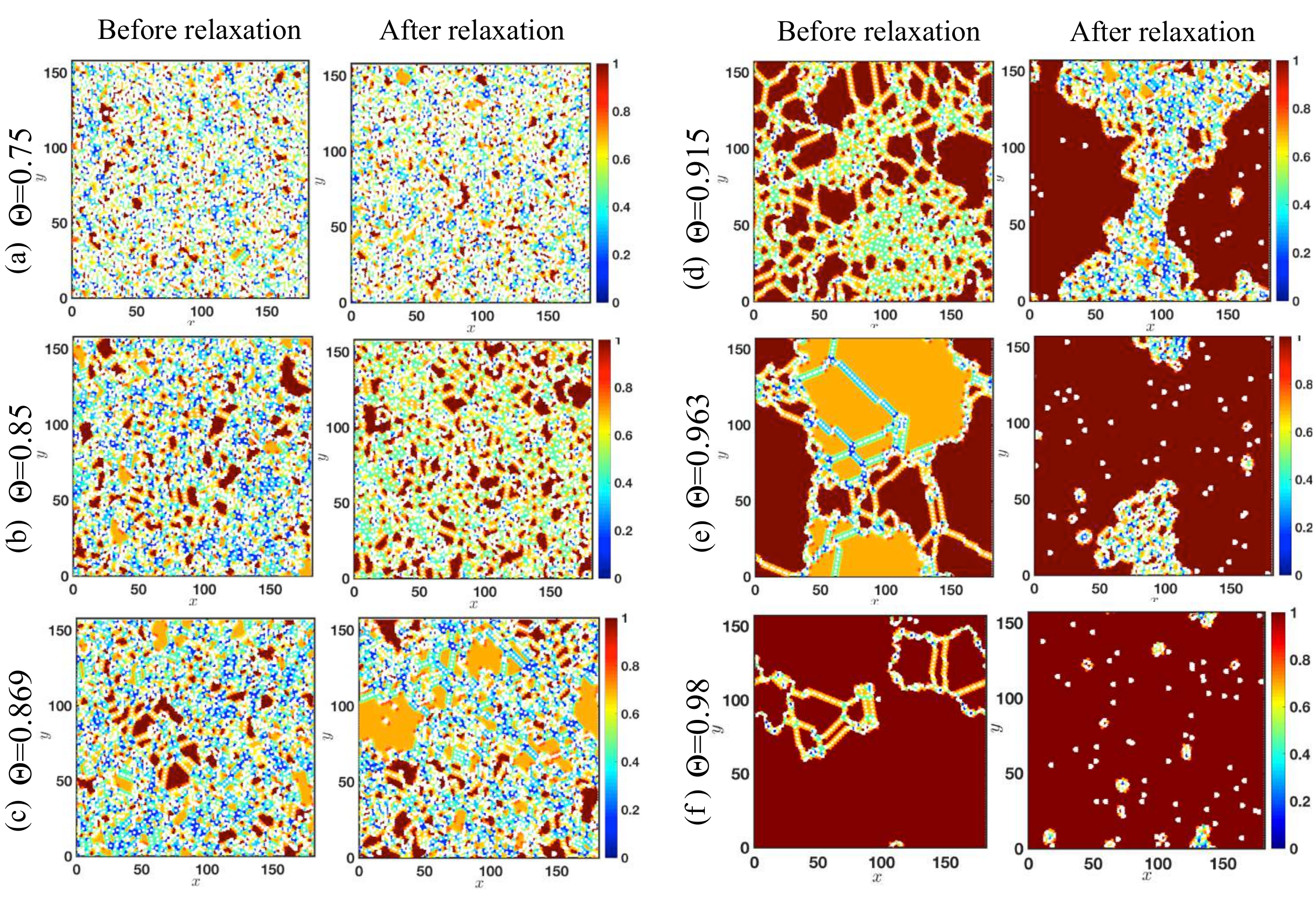}
\caption{ Local bond orientation order based on relaxation method at low surface diffusion $D=0.01$ and different surface coverage: (a) $\theta=0.75$, (b) $\theta=0.85$, (c) $\theta=0.869$, (d) $\theta=0.915$, (e) $\theta=0.963$, (f) $\theta=0.98$. Before relaxation refer to the configuration obtained from adsorption method at $D=0.01$ in the first step of relaxation method , and after relaxation refer to the equilibrium configuration.}
 \label{fig8}
 \end{figure*}
Finite size effects are another important key issue in finding the equation of state in lattice simulations. Figure.~\ref{fig4} presents the variation in adsorption rate for triangular lattices covering $7$ sites of sizes $105$ to $266$. For surface diffusion of $D=1$, the same blocking function is obtained from both adsorption and desorption methods at low surface coverage as presented in Figure.~\ref{fig4}(a). Initially, in the vicinity of the phase transition, by increasing the lattice size dimension, the adsorption rate increases for both the adsorption and desorption method as illustrated in Figure.~\ref{fig4}(b). Exactly before all of the curves overlap around full coverage, the adsorption rate decreases by increasing the lattice dimension for both methods. The results in Figure.~\ref{fig4}(b) show that the adsorption rate is significantly different for $d=105$ in comparison to $d=196$ and $d=266$. As a result $d=105$ is not a reliable lattice for finding the equation of state. Later on we will discuss this case and compare it with $d=196$ in Figure.~\ref{fig5}(b) and Figure.~\ref{fig6} only to show the importance of using a large enough system. One of the advantages of using the $RSAD$ method is we know how big our system should be to ensure that the results are at the same time accurate and computationally less expensive. \par
Although an exact solution of this model does not exist in the literature, meaning there is no certain agreement upon the equation of state and its phase transition, we can compare our results with the analytic calculation of Orban and Bellmans \cite{orban1968phase} for $d=196$ and $D=80$ in Figure.~\ref{fig5}(a). This paper used two different methods, the matrix method, based on the sequence of exact solutions for lattices of infinite length and increasing finite width, and the series expansion method, based on knowing the final structure at close packing and constructing the density and activity series to find the surface pressure. A first order transition was found for both the matrix and the series expansion methods, where the transition occurs at surface pressure of $4.5-5$ for surface coverage of ${\rm{0.81}} \pm {\rm{3}}$ and ${\rm{0.98}} \pm {\rm{1}}$ for fluid and solid regimes respectively. This system is also studied by Chestnut \cite{orban1968phase} using a Monte Carlo technique, but the study was limited to surface coverages lower than $0.75$, where no phase transition was found.\par
As illustrated in Figure.~\ref{fig5}(a), at low surface coverage there is no difference between the reported equations of state. However, in the lower part of the phase transition zone, our equation of states shows lower surface pressure than Orban and Bellman's series expansion method (low and high density) but higher surface pressure than their matrix method. On the contrary, in the upper part of the phase transition region, the matrix method shows higher surface pressure than what we found, but the series expansion method overlaps with both of our methods. The thermodynamically-stable surface pressure loop is observed in our simulation results as is reported extensively in the literature due to finite size effects \cite{bernard2011two,engel2013hard,mayer1965interfacial,schrader2009simulation}. Surface pressure can create a thermodynamically stable loop at equilibrium for a finite size system, but this loop will disappear at infinite size and the coexistence zone will be flat. Creation of the flat pressure is visible in Figure.~\ref{fig5}(b) for $d=196$ around a surface pressure of ${\rm{4.5}} \pm {\rm{0.05}}$. Moreover, based on the lower part of the phase transition in Figure.~\ref{fig4}(b), the adsorption rate tends to increase by increasing the lattice dimension, which means less surface pressure will be expected for a larger lattice size. Conversely, the adsorption rate tends to decrease for both the adsorption and desorption methods in the upper part of the phase transition, which means higher surface pressure will be expected for a larger lattice size. This behavior indicates the tendency of the system towards flatness at an infinite size. From the equality shown in the hatched area of Figure.~\ref{fig5}(b), the horizontal surface pressure is obtained from the Maxwell construction, where overlapping of this construction line with the flat region of the equation of state confirms the tendency of the system to be flat at infinite size. Figure.~\ref{fig5}(b) shows that by increasing the lattice dimension, surface pressure decreases in the lower part of the phase transition and increases in the upper part of phase transition. Bernard et al \cite{bernard2011two,engel2013hard} reported the same trend for the equation of state of the melting transition of hard disks by increasing the number of particles where they reported a two step transition; first order liquid-hexatic and continuous hexatic-solid transition for their system.\par
The phase transition zone is studied in Figure.~\ref{fig6} based on the derivative of surface pressure with respect to surface coverage in the adsorption method. One can see from Figure.~\ref{fig6}(b) that for an insufficiently large system, the first order liquid-solid phase transition is obtained for $d=105$ and $D=80$; however, a different phase transition is obtained for $d=196$ and $D=80$ in Figure.~\ref{fig6}(a), which indicates the importance of using a sufficiently large system. Phase transition peaks obtained from Figure.~\ref{fig6}(a) for $d=196$, $D=80$ are analyzed through a bond orientation correlation function, g6(r), in Figure 7 based on the relaxation method for each individual configuration. For additional insight into the phase transition region, each of these configurations is visualized via the local bond orientation order function $\Psi ({r})$, based on the relaxation method, in Figure.~\ref{fig8}. The first peak in the phase transition curve corresponds to surface coverage of $0.826$. Since we find the same exponential decay of ${g_6}(r)$ before and after relaxation at surface coverage of $0.75$ in Figure.~\ref{fig7}(a) suggests that the system is in a pure liquid regime below a surface coverage of $0.826$. Furthermore, the homogeneous distribution of local bond orientation order over the surface in Figure.~\ref{fig8}(a) confirms the presence of a liquid phase below surface coverage of $0.826$. In Figure.~\ref{fig6} we see that lattice size does not have any impact on the liquid phase since the same behavior from liquid regime is obtained for both $d=105$ and $d=196$ through fractional surface coverages up to $0.826$. \par
\begin{figure*}[!htb]
\centering
\includegraphics[width=1\textwidth]{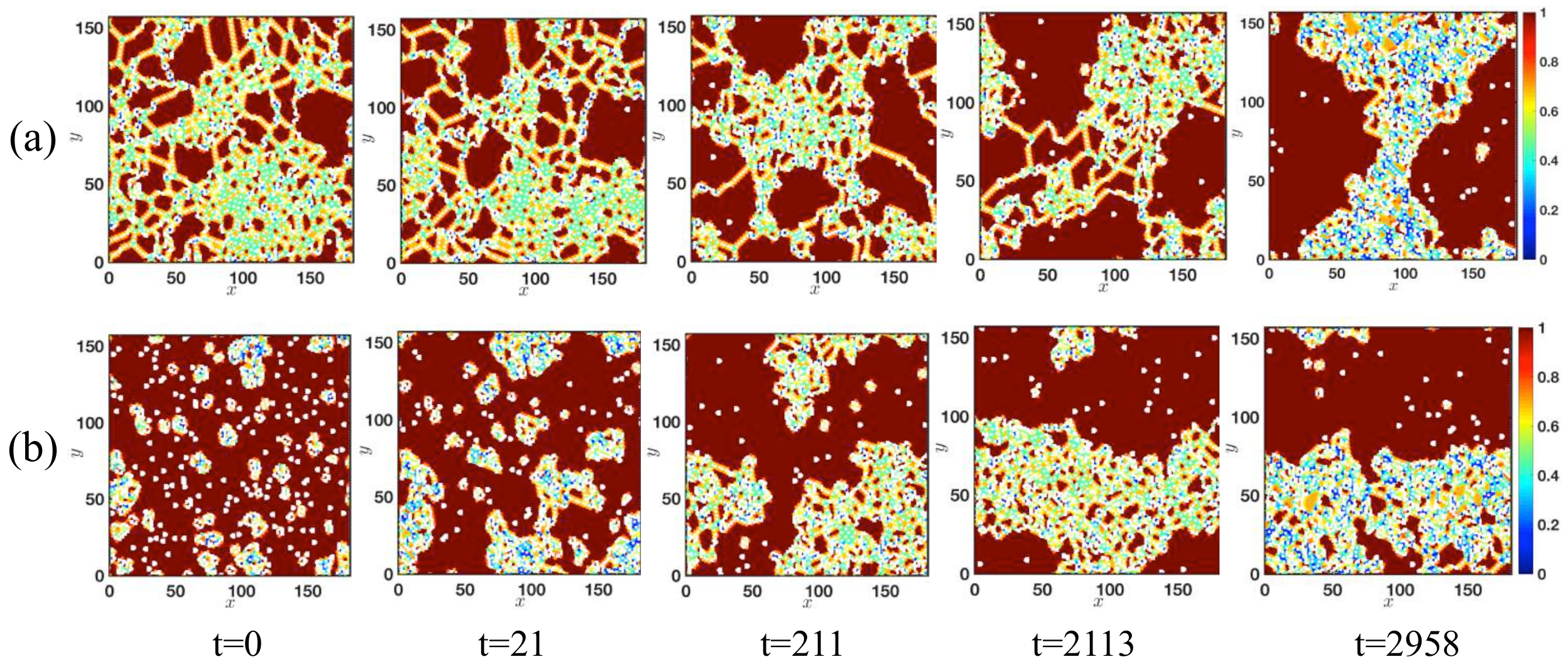}
\caption{Local bond orientation order for surface coverage of 0.915 with D=0.01 and d=196, (a) for adsorption method over time, (b) for desorption method over time.}
 \label{fig9}
 \end{figure*}
The transition between positive and negative slopes occurs at surface coverage of $0.864$, located between the first and second peaks, so we analyze surface coverages of $0.85$ and $0.869$ to see the behavior of the system below and above the transition value. For surface coverage of $0.85$, ${g_6}(r)$ decays exponentially at the same rate before and after relaxation in Figure.~\ref{fig7}(b); more peaks were observed after relaxation, which is an indication of the creation of a more ordered phase in the system. Creation of a more ordered phase is also confirmed by the local bond orientation order seen in Figure.~\ref{fig8}(b), indicated by more red spread through the surface. For surface coverage of $0.869$, ${g_6}(r)$ still decays exponentially, but at different rates before and after relaxation as illustrated in Figure.~\ref{fig7} (c-1) and (c-2), respectively. The local bond orientation order in Figure.~\ref{fig8}(c) shows that ordered particles prefer to stick to each other. The third peak in Figure.~\ref{fig6} corresponds to surface coverage of $0.915$. At this coverage the system decays exponentially before relaxation as presented in Figure.~\ref{fig7}(d-1) and Figure.~\ref{fig8}(d) shows that the cluster of ordered phase spreads through the surface and indicates a glassy state. After relaxation, ${g_6}(r)$ decays with a power law with $\theta=0.25$ as illustrated in Figure.~\ref{fig7}(d-2) , which is an indication of a hexatic phase. The local bond orientation order shown in Figure.~\ref{fig8}(d) reveals that at this coverage all the particles tend to stick together and they are between the mobile particles, as was also reported in reference \cite{hou2019phase}. The tendency of ordered particles to stick together causes the probability of success to increase in Figure.~\ref{fig2}(b) due to maximizing the available free surface for accepting the new incoming particles. By increasing the surface coverage from 0.915 to 0.963, $\eta$ decreases from $0.25$ to $0.08$ after relaxation as illustrated in Figure.~\ref{fig7}(e-2), which indicates that the fourth peak in Figure.~\ref{fig6} corresponds to a first order transition from hexatic phase to solid phase in equilibrium state.  \par
To further illustrate the relaxation dynamics toward an equilibrium state, the local bond orientation order parameter of particles at surface coverage of $0.915$ was tracked over time for two different initial configurations obtained from adsorption and desorption methods at very low surface diffusion ($D=0.01$, $d=196$). Figure.~\ref{fig9}(a) indicates that the density should be increased very slowly when we start from an empty lattice, or else the system will be locked into a glassy configuration \cite{alder1962phase}. A glassy state is reported in reference \cite{hou2019phase} during rapid compression of a system composed of spherical particles. For the desorption method, Figure.~\ref{fig9}(b), the system is initially ordered, so the density should be decreased very slowly or else the system will be metastable \cite{alder1962phase}. As time passes in the equilibrium state, phase separation occurs, where ordered particles tends to cluster and create more space for further adsorption.  \par

\section{Conclusion}

In this paper we studied the phase behavior of a hard-core molecules with third neighbor exclusions in a triangular lattice with $RSAD$ simulations, in order to derive the equation of state of a two-dimensional hard-core particle based on kinetic arguments and the Gibbs adsorption isotherm. We compared our results with Orban and Bellman \cite{orban1968phase}, who used a matrix method and a series expansion methods in low and high densities, and found only partial agreement. Our results show that the system is in a pure liquid regime below surface coverage of $0.826$. Increasing the surface coverage will create a more ordered phase where a first order liquid-hexatic phase transition occurs between surface coverage of $0.877$ and $0.915$. At surface coverage of $0.915$, ${g_6}(r)$ decays algebraically after relaxation with a power-law exponent $\eta=0.25$, which is an indication of a hexatic phase. Our simulation results reveal that as the surface coverage increases, after relaxation, the surface particles tend to form tightly-packed clusters in like-oriented domains, while the remaining mobile particles have more random orientations. By increasing the surface coverage above $0.915$, $\eta$ decreases from $0.25$ toward zero after relaxation, where the system undergoes a first order transition from a hexatic phase to a solid phase. \par
One of the advantages of using the $RSAD$ model is being able to locate the equilibrium state, which assures us that adequate thermalization and finite size are being used to reach the equilibrium state.  Moreover, subtle details of the clustering structure, through direct visualization of the system using the relaxation method at any fractional surface coverage, provides insight regarding coexistence regions and phase transitions.\par

\section*{Acknowledgments}
This research was supported by the National Science Foundation under grant $No. 1743794$, PIRE: Investigation of Multi-Scale, Multi-Phase Phenomena in Complex Fluids for the Energy Industries.

\end{document}